\documentclass[12pt]{article} 
\usepackage[hyperfootnotes=false]{hyperref}
\usepackage{amsmath}
\usepackage{amssymb}
\usepackage{graphicx}
\usepackage{graphbox}
\usepackage[dvipsnames]{xcolor}
\usepackage{float}
\newlength{\abstractwidth}
\usepackage{comment}
\usepackage{mciteplus} 
\usepackage{subcaption}

  \newcommand{\be}{\begin{equation}}
  \newcommand{\bea}{\begin{eqnarray}}
  \newcommand{\eea}{\end{eqnarray}}
  \newcommand{\ee}{\end{equation}}

	\newcommand{\eqn}[1]{\begin{equation}\begin{split} #1 \end{split}\end{equation}}
	
	\newcommand{\lp}{\left (}
	\newcommand{\rp}{\right )}
	
	\newcommand{\rb}{\right]}
	
	\newcommand{\RA}{\Rightarrow}

	\newcommand{\pa}[2]{\frac{\partial #1}{\partial #2}}
	\newcommand{\pd}{\partial}
	
	\newcommand{\inv}{^{-1}}

	\newcommand{\hf}{\frac{1}{2}}

	\newcommand{\ket}[1]{\left| #1 \right \rangle}
	
	\newcommand{\ev}[1]{\left \langle #1 \right \rangle}

\def\bsub{ \begin{subequations}
\begin{empheq}[box=\widefbox]{align}  }
\def\esub{ \end{empheq}
\end{subequations}}

\def\bn{\bigskip \noindent}

  \def\r{\bf{r}}

  \def\r{\rho}

\def\eq{&=&}

\def\la{\langle}

\def\simleq{\; \raise0.3ex\hbox{$<$\kern-0.75em
\raise-1.1ex\hbox{$\sim$}}\; }
\def\simgeq{\; \raise0.3ex\hbox{$>$\kern-0.75em
\raise-1.1ex\hbox{$\sim$}}\; }

\def\bi{\begin{itemize}}
\def\ei{\end{itemize}}

\def\CC{{\cal{C}}}

\def\CL{{\cal{L}}}

\def\t{\tau}

\def\uk{u_\text{kick}}

\def\la{\label}

\def\32{{3 \over 2 } }

\def\ut{\tilde{u}}

  \def\ba{\begin{eqnarray}}
  \def\ea{\end{eqnarray}}

 \def\simleq{\; \raise0.3ex\hbox{$<$\kern-0.75em
      \raise-1.1ex\hbox{$\sim$}}\; }
 \def\simgeq{\; \raise0.3ex\hbox{$>$\kern-0.75em
	\raise-1.1ex\hbox{$\sim$}}\; }

\def\nref#1{(\ref{#1})}

\begin{document}

\begin{titlepage}
 % \rightline{}
  \bigskip

  \bigskip\bigskip

  \bigskip

\begin{center}
 
\centerline
{\Large \bf {Complexity Geometry and Schwarzian Dynamics}}
\bigskip 
%{or\\}
%\bigskip
%{\Large \bf {Complexity, the Schwarzian, and Newton }}
 %\bigskip

 \bigskip
%\centerline
{\Large \bf { }} 
    \bigskip
\bigskip
\end{center}

  \begin{center}

	  \bf {Henry W. Lin$^{1,3}$ and Leonard Susskind$^{2,3}$ }
  \bigskip \rm
  
\bigskip
 $^1$Jadwin Hall, Princeton University, Princeton, NJ 08540, USA\\

 \rm 
 \bigskip
 $^2$Stanford Institute for Theoretical Physics and Department of Physics, Stanford University, Stanford, CA 94305, USA\\
\bigskip
 $^3$ Google, Mountain View, CA 94043, USA 
\rm
 \bigskip

 % \bf {Write authors  }
  \bigskip \rm
\bigskip
 
 %   Institute for Advanced Study,  Princeton, NJ 08540, USA  \\
\rm

\bigskip
\bigskip

% \vspace{2cm}
  \end{center}

 \bigskip\bigskip
  \begin{abstract}
	A celebrated feature of SYK-like models is that at low energies, their dynamics reduces to that of a single variable. In many setups, this ``Schwarzian'' variable can be interpreted as the extremal volume of the dual black hole, and the resulting dynamics is simply that of a 1D Newtonian particle in an exponential potential. 
	On the complexity side, geodesics on a simplified version of Nielsen's complexity geometry also behave like a 1D particle in a potential given by the angular momentum barrier.
	The agreement between the effective actions of volume and complexity succinctly summarizes various strands of evidence that complexity is closely related to the dynamics of black holes.

	  % Nearly AdS$_2$ gravity, the physical degrees of freedom is just the extremal distance.
	  %	In Nearly AdS$_2$ setups, the 
	 %effective degrees of freedom is just the volume.
	  %In many setups, the Schwarzian degree of freedom in the SYK model and in Jackiw-Teitelboim gravity behaves in many situations like a non-relativistic particle in an effective potential. Assuming complexity = volume, this feature is naturally explained by the Newtonian character of the geodesic equation on a smooth complexity geometry. 
	  %If we assume complexity = volume, this feature is naturally explained by complexity geometry.

% \medskip
%  \noindent
  \end{abstract}
\bigskip \bigskip \bigskip

\vspace{1cm}

\vspace{2cm}

  \end{titlepage}

  %  \starttext \baselineskip=17.63pt \setcounter{footnote}{0}
   \tableofcontents

 % \sc
   %\section*{STUFF TO DO}
   %\begin{enumerate}
%	   \item check all formulas that need to be checked
%	   \item finish conclusion
%	   \item size section needs to be streamlined
	   %\item Finalize appendices. Straighten out some conventions.
 %  \end{enumerate}

\section{Introduction} 

There are reasons to believe that the evolution of quantum complexity---a purely information theoretic quantity---is guided by dynamical equations which follow from an action principle; and that those equations are closely related to the dynamical equations of gravity. Let us list a few of those reasons:

\begin{enumerate}

	\item The growth of the interior of a black hole---a process governed by Einstein’s equations---is believed to be dual to the growth of the quantum complexity of the holographic state of the black hole system \cite{Susskind:2014rva, Susskind:2014moa, Stanford:2014jda}.

	\item  The holographic dual of gravitational attraction is the tendency for operator growth  \cite{Roberts:2018mnp,Susskind:2018tei,Brown:2018kvn,Qi:2018bje,Lin:2019qwu}. The correspondence between the two has been precisely formulated for the SYK model \cite{Sachdev:1992fk,KitaevTalks,Maldacena:2016hyu} and its gravitational dual at low energies, Jackiw-Teitelboim gravity \cite{Almheiri:2014cka,Jensen:2016pah,Maldacena:2016upp,Engelsoy:2016xyb}. Operator growth or scrambling, is the early time manifestation of the growth of complexity. 

%(For other toy models of complexity geometry, see \cite{Lin:2018cbk, Brown:2019whu}.)

%\item  The Complexity-Action correspondence CITE suggests a connection between gravitational dynamics and complexity.

	\item The Nielsen geometric approach to complexity \cite{dowling2008geometry} can be formulated as a non-relativistic particle moving on a group manifold with a somewhat unusual metric. Even the lower curvature version of the complexity geometry defined in \cite{Brown:2017jil} is difficult to analyze at large distances, but the main points of complexity geometry have been illustrated in a simple ``toy complexity geometry" (TCG) based on the two-dimensional Poincare disc \cite{Brown:2016wib}. \end{enumerate}

In this paper we will apply the TCG to the SYK system and its bulk dual, JT gravity. In a  somewhat surprising and nontrivial way the complexity geometry approach reproduces the gravitational dynamics of the boundary Schwarzian theory. This agreement neatly summarizes various tests of the complexity = volume conjecture.
We regard this agreement as additional evidence that gravitational dynamics reflects the general principles of quantum complexity.

\def\cj{\mathcal{J}}
We will assume that the reader is familiar with the general ideas of complexity geometry, and briefly review the TCG of \cite{Brown:2016wib}.
We will work with SYK units for most of this paper. The SYK model has three parameters: the number of Majorana fermions $N$, the dimensionless coupling $\beta\cj$, and $q$ the number of fermions in the interaction.
The semi-classical Schwarzian limit corresponds to $1 \ll \beta \cj \ll N $.

\section{Effective actions in Complexity Geometry}
\subsection{Toy Complexity Geometry}

Ideally we would like to start with a microscopic definition of complexity, and then derive its various properties. At present, such an approach is beyond our capabilities. TCG is motivated by a more phenomenological approach, where we write down a model that is consistent with various features we expect complexity to have.

The TCG of \cite{Brown:2016wib} is a high-genus compactification of the Poincare disc. If, as in this paper, we are not interested in exponentially long time scales, we can ignore the compactification and simply work with the Poincare disc. The metric is
\bea 
dl^2 = R^2 \lp d\r^2 + \sinh^2{\r} \  d\theta^2\rp \la{poincare}
\eea
where $R$ is the curvature scale of the disk and $\rho$ is a dimensionless radial coordinate. The distance to the origin, which we interpret as the state complexity, is $R \rho$.

Now consider a non-relativistic particle moving on this geometry. Two important parameters of the model are the speed of the particle and the curvature scale $R$. To determine the speed of the particle, we demand that the complexity should grow at a rate $dC/d \tau = N$. Here the circuit time is $\tau = 2\pi u/\beta$, where $u$ is the physical time of the quantum mechanical system\footnote{We are following the somewhat pathological practices of Princeton SYK (pronounced Psych) workers.}. In a holographic setup $\tau$ is Rindler time and $u$ is the asymptotic time on the boundary. Identifying the distance with complexity sets $v = ds/du = 2\pi N/\beta$.

To determine $R$, we consider two particle trajectories with speed $v$ emanating from the same point, but headed in slightly different angular directions. It is easy to see that the distance between them grows like 
\eqn{\delta L \propto R e^{v \tau/R} \delta \theta}
Since the Lypapunov exponent $\lambda = v/ R  = 1$, this sets $R = N$.
\def\ee{\mathfrak{J}}
In addition to these physical parameters, there is an unphysical parameter, the mass of the fictitious particle. The choice of mass is entirely conventional since changing it just amounts to multiplying the action by a constant, which does nothing in classical physics. We will choose a convention where the kinetic energy scales with the system size $E = \hf M v^2 \propto N /\beta^2$, or $M = 1/(N\ee)$, where $\ee$ is some arbitrary constant with units of inverse length, chosen so that energy has the correct dimensions. With these conventions, the action is
\def\shr{\sinh^2\!\r }
\eqn{S = {N \over 2 \ee}  \int \dot{\r}^2 +\shr \, \dot{\theta}^2 \, du }
The particle has a conserved charge $J =  \shr \, \dot{\theta} $. We can use this conserved charge to dimensionally reduce the $\theta$ coordinate:
\eqn{S = {N \over 2 \ee}  \int \dot{\r}^2 -  {J^2 \over  \shr } du \approx {N \over  \ee}  \int \hf \dot{\r}^2 -  2 {J^2 e^{-2 \rho} } du  \la{cgea}}
In approximating the sinh as exponential, we are assuming that complexity is never small. This will be appropriate for low energy states. We end up with a particle in an exponential potential.
\begin{figure}[H]
\begin{center}
\includegraphics[scale=.25]{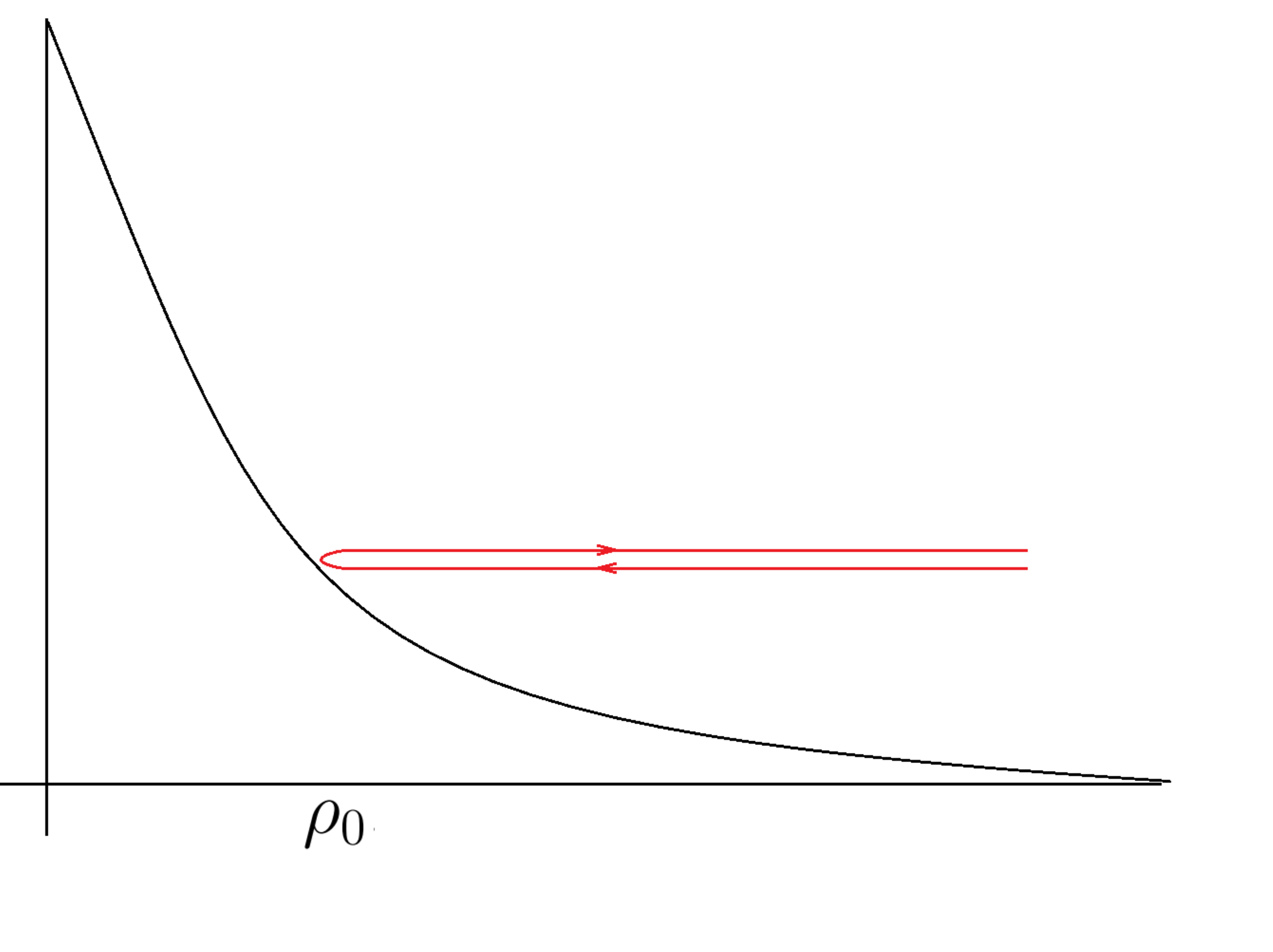}
%\caption{     }
\label{bounce}
\end{center}
\end{figure}
The particle comes in from infinity, rolls up the potential, and then turns around at some $\rho_0$.
This exponential potential is of course nothing but the angular momentum barrier which ensures that geodesics are circles intersecting the boundary of the disk at right angles.
Since the asymptotic velocity (or equivalently the kinetic energy) of the particle is fixed, the parameter that controls the distance to closest approach $\rho_0$ is simply $J$.

%Then if we convert to boundary time $u$,
%\eqn{S = {N \over 2}  \int {\beta \over 2 \pi}  {\r'}^2 +  {J^2 \over \sinh^2 \rho} d u }

%This is a reasonable convention because if we kick the particle by changing its velocity $\delta v \sim 1$ we get a change in energy of order $\sim 1$ in this convention. $L^2
%A change in velocity $\delta v \sim 1$ corresponds to 

\begin{figure}[h]
\begin{center}
	\includegraphics[scale=0.4]{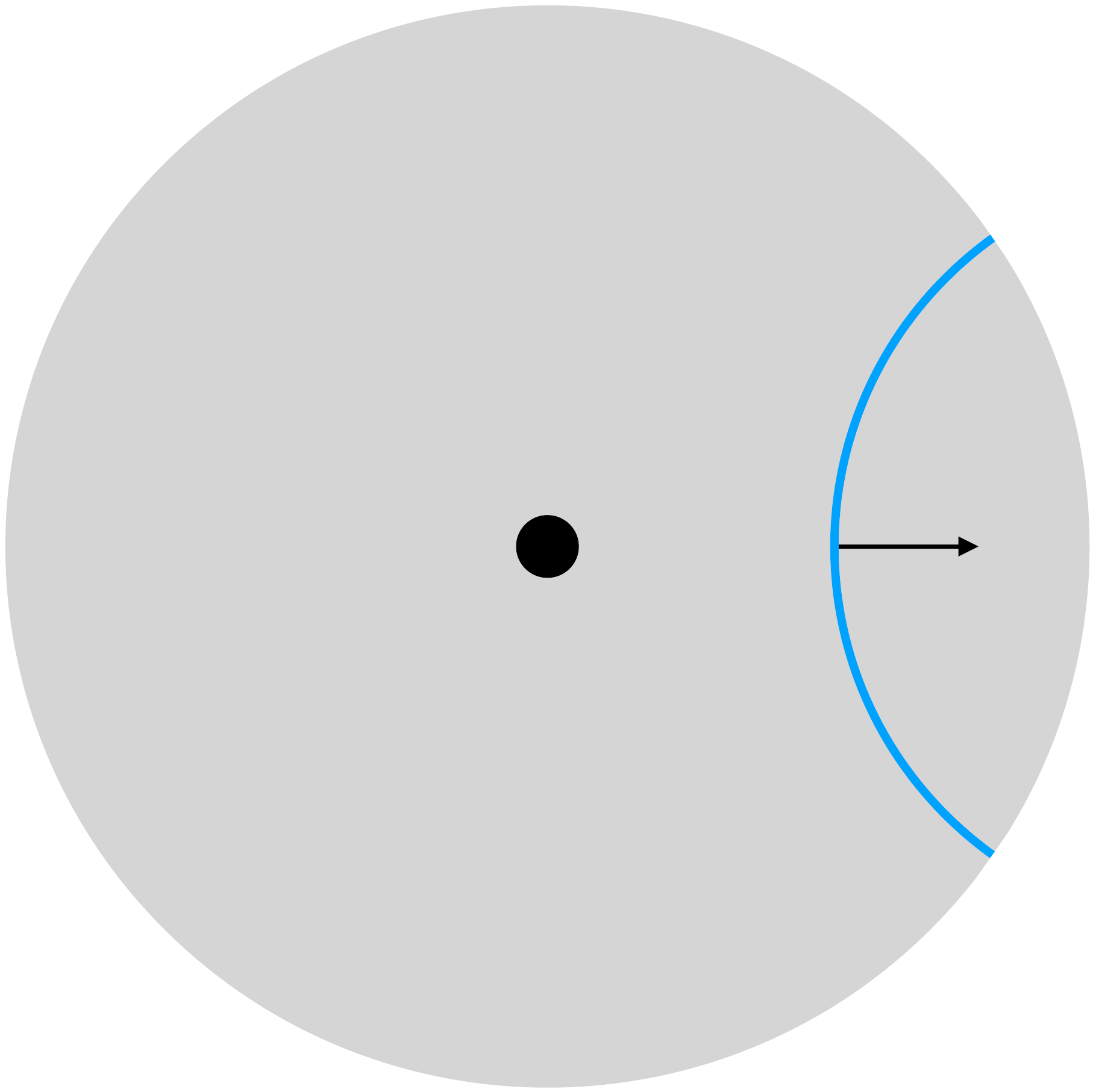} 	
	\caption{A particle on the hyperbolic disk. We indicate the centrifugal force (which is always repulsive). }
	\label{hyperbolic}
\end{center}
\end{figure}

\begin{figure}[h]
\begin{center}
	
\includegraphics[align = c, scale=0.35]{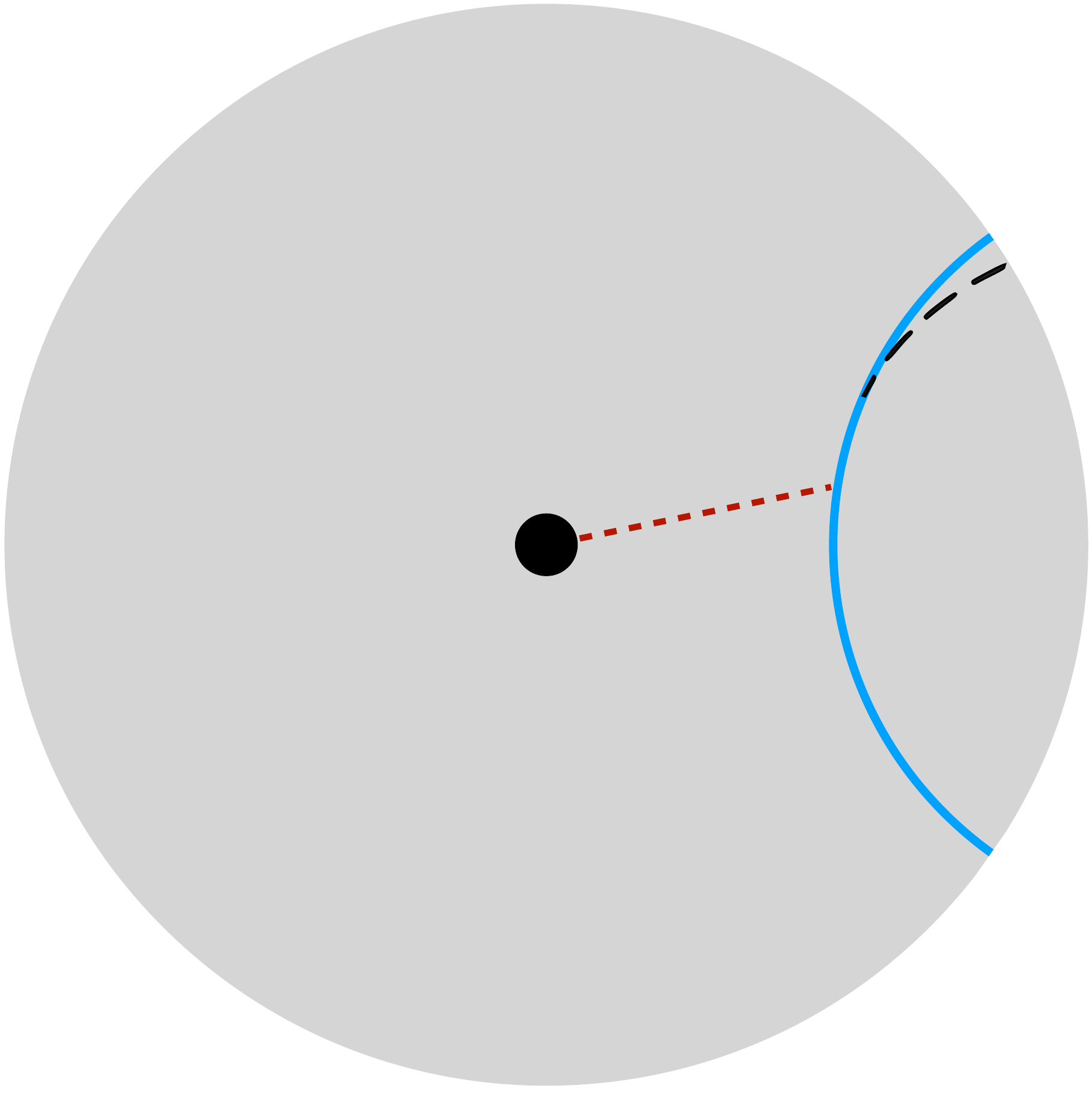}
\hspace*{.3in}
\includegraphics[align = c, scale=0.35]{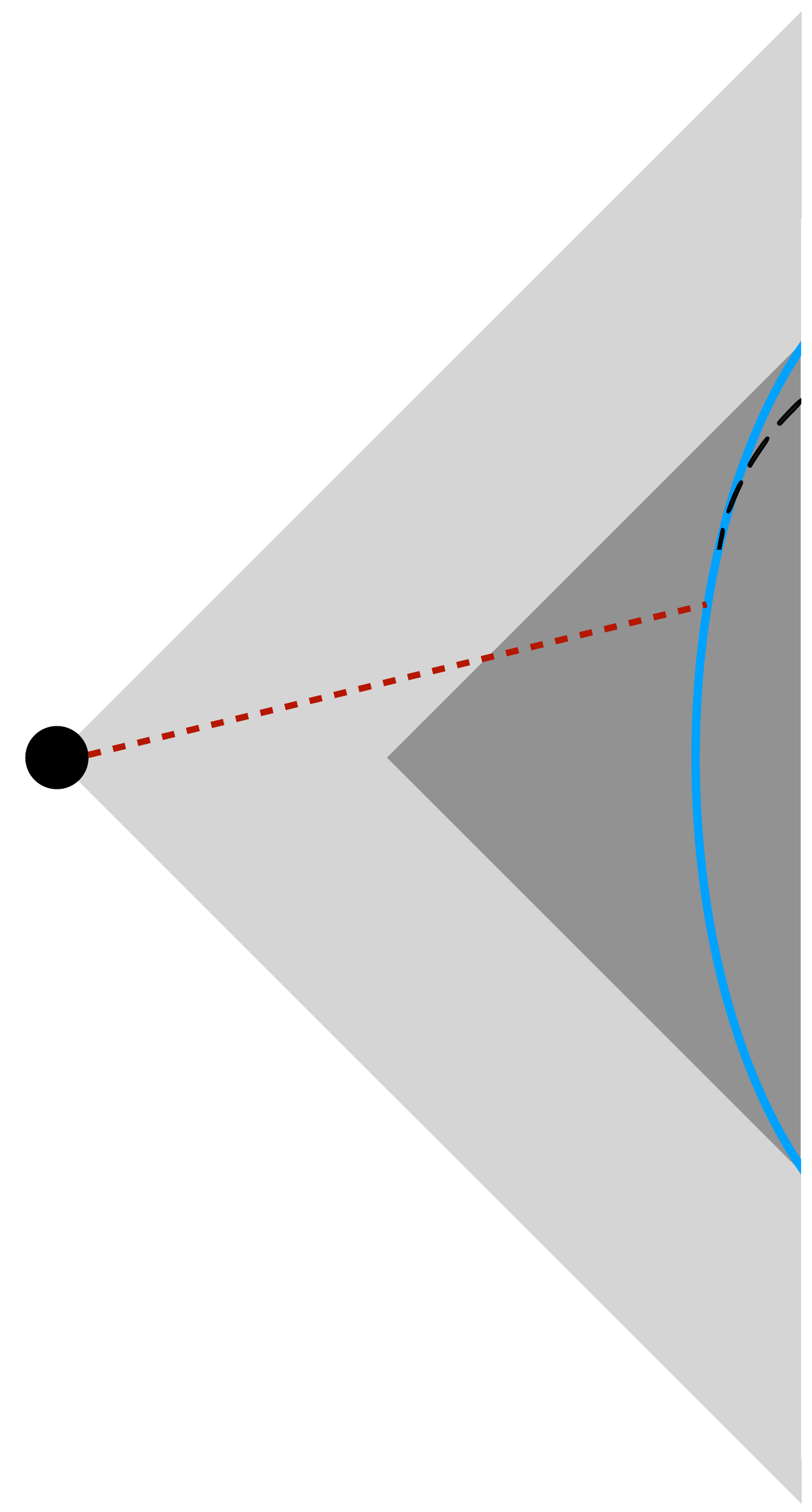}
\caption{Summary of our quantitative match. We consider the length of the a 1-sided black hole (with an end of the world brane), potentially with a kick dictated by SL(2) charge conservation. This matches the complexity as calculated in the toy model. The change in trajectory is due to the additional angular momentum when a perturbation to the particle is applied. The length of the dashed red lines in both figures are to be identified.}	
	\label{pure-state}
\end{center}
\end{figure}

The effective action however tells us a bit more than just the geodesic solutions. In particular, we can consider what happens when we kick the particle. At the time of the kick, the angular momentum is increased slightly. In our conventions, $J$ is of order 1, so a singe-gate perturbation should should correspond to a shift in the angular momentum of order $1/N$. However, due to the butterfly effect, the distance between the counterfactual trajectory of the unperturbed particle and the perturbed particle grows exponentially, see Figure \ref{pure-state}. 
We will have more to say about this in section \ref{sec-switchback}.

\def\tcg{\cite{Brown:2016wib}}
\subsection{Effective action on positively curved space}

\begin{figure}[h]
\begin{center}
	\includegraphics[scale=0.4]{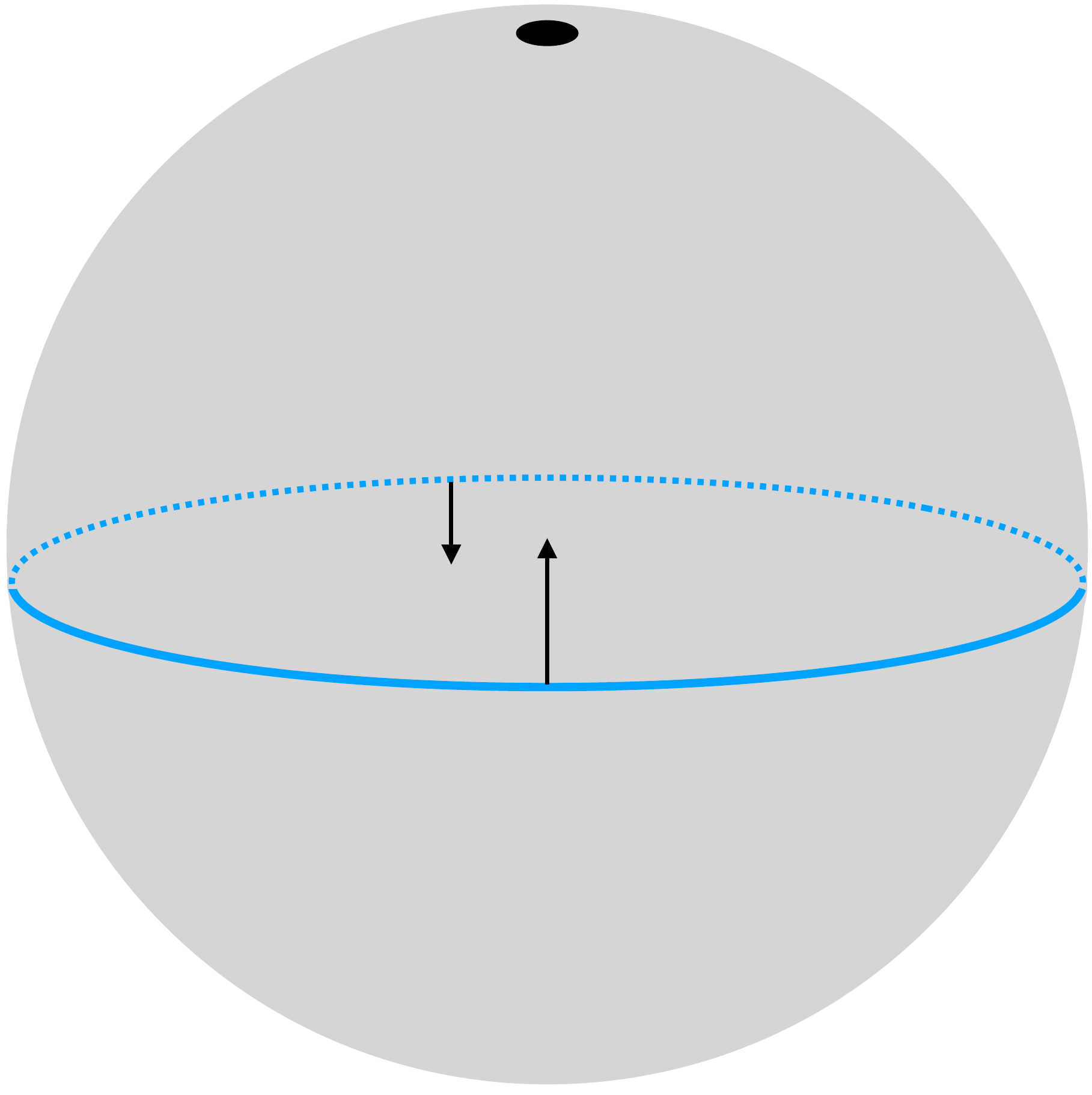} 	
	\caption{A particle on a positively curved space. The centrifugal force is sometimes attractive and sometimes repulsive.}
\label{sphere}
\end{center}
\end{figure}

The model discussed in the previous section is suitable for maximally chaotic quantum systems. 
Indeed, we used the Lyapunov exponent to calibrate the negative curvature of the model. 
We now consider what complexity geometry looks like in the opposite regime, for systems that are non-chaotic and integrable (at least in some approximation). 
The main application we have in mind are systems like the harmonic oscillator, where a large class of states return to themselves after some period of time $T \sim O(1)$, which is much shorter than the quantum recurrence time $\sim e^{e^N}$ we would have expected in the chaotic case. 

We argue that the main qualitative feature of the complexity geometry is positive curvature. Positive curvature means that geodesics tend to converge instead of diverge. This is necessary in order for states to recur on a small timescale.
Just to be more concrete, we can consider the simplest example of a positively curved space, the $2$-sphere:
\eqn{ds^2 = d\theta^2 + \sin^2 \theta d\phi^2.
}
If we take the origin to be the North pole $\theta = 0$, then the distance to the origin is just $\theta$. We can describe the dynamics of $\theta$ by an effective potential
\eqn{L &= \hf \dot{\theta}^2 - V,\\
V &= {J^2  \over 2\sin^2 \theta} \approx {J^2 \over 2} \lp 1+ \delta \theta^2\rp, \quad \delta \theta = \theta - \pi/2 }
The main point is that in a curved space, the effective potential becomes stable as opposed to repulsive. The stable point corresponds to a particle wrapping the equator of the sphere. The time it takes the particle to wrap around the circle is the period $T$ that it takes for a state to return to itself.

In section \ref{coupledHam}, we will consider the coupled SYK system \cite{Maldacena:2018lmt} which is dual to global AdS$_2$. Since global AdS$_2$ has no horizon, the system is not chaotic. In this model, there are a class of low energy states where the complexity is constant. From the complexity geometry point of view, these are like states which are rolling around the equator of the sphere. Perturbing such states leads to oscillations in the complexity, which also matches bulk expectations.

\section{Effective dynamics of bulk distance}
In this section, we discuss the effective dynamics of the distance (conjectured to be dual to complexity). We consider a smorgasbord of two-sided and one-sided setups; the upshot is that in each situation there is a close and sometimes exact match between the effective actions derived from JT gravity, and the ones derived from TCG.

\subsection{2-sided case}
\def\dlr{\ell_{LR}}
\def\vf{\varphi}
We will be interested in Lorentzian JT gravity + matter in the Schwarzian limit. In this limit, the effective dynamics is governed by time reparameterizations on the left and right boundaries $T_l(u)$ and $T_r(u)$. Here $u$ means evolution by $H_l + H_r$.
We will be further interested in left-right symmetric configurations, in a gauge where $T_l = T_r = T(u)$. As discussed in the Appendices, we can introduce a variable $-2\vf$, which is a regulated geodesic length between the two sides:
\eqn{\vf = - {\dlr \over 2L_\text{AdS} } + \text{const}.} 
Furthermore, as discussed in the Appendix, the 2-sided Schwarzian action gives a very simple non-relativistic effective action for this length variable:
\def\tu{\tilde{u}}
\eqn{S = N \int d\tilde{u} \, {\vf'}^2 - V(\vf), \quad V = e^{2\vf} + {E\over N } e^\vf. }
Here and below, we use the dimensionless boundary time $\tilde{u} = u \cj/\alpha_S$; equivalently we measure time in units where the coupling $\cj/\alpha_s$ is set to 1. All bulk lengths will be measured in AdS units. Primes denote derivatives with respect to $\tilde{u}$. $E$ is the global energy of bulk matter. We can interpret $e^\varphi = T'(\tilde{u})$ as a conversion factor between the global energy $\pd_T$ and the boundary energy $\pd_u$.
For the thermofield at any temperature and time $\beta + i u$ the matter energy vanishes $E=0$. More generally, we can start with the TFD, evolve forward and/or backwards in time, and apply symmetric kicks to the boundary. The entire effect of these shockwaves on the distance variable $\vf$ is accounted for by changing the value of $E$ each time a shockwave is inserted. Both $\vf$ and $\vf'$ are continuous across the shock.

%A word of caution: if we excite $O(1)$ fermions to create a shockwave, this changes the boundary energy by an amount that is $\sim \cj$. However, the global energy $E$ changes by an amount that is time dependent. For example, if we perturb the TFD at some Lorentzian time $u$, this leads to a shockwave of global energy $E \sim \cj e^{-\vf} \sim {\beta \cj  \cosh (u/\beta) }$.

It might seem strange that the effective potential of the boundary particle changes when matter is inserted. This can be explained in the 4D Near-Extremal black hole picture \cite{Susskind:2019ddc}. As a particle falls down the long throat, it experiences an effective S-wave potential. The top of the potential is defined to be the JT boundary. If the potential pushes on a matter particle, it must also lead to a backreaction on the boundary. The force on the particle due to the potential decays exponentially with proper distance, which means that there is also an exponentially decaying force on the boundary, in agreement with our analysis.

In order to match with the effective action \nref{cgea}, we should identify
\eqn{4 J^2 = 1, \quad \phi = - \rho + \text{const}. \la{match} }
We could also choose $\ee$ to match the numerical values of the action, but as we emphasized before, there is no physical content.

We have so far discussed the effective action for the distance variable $\varphi$. We could alternatively work in the Hamiltonian formalism of JT gravity, see \cite{Harlow:2018tqv}. The resulting Hamiltonian\footnote{See v2 of Harlow-Jafferis \cite{Harlow:2018tqv}, section 2.2, for a correction that is important for our purposes.} is of the form $H = \cj/N\alpha_s \lp P^2 + 4 e^{-\ell} \rp $, which is consistent with our action. Note that the numerical coefficient in front of the exponential potential depends on how we regulate the length. 

%, but a small error in their expression for the length led to a complicated expression for the Hamiltonian. %In Appendix B, we show that with the corrected expressions, we get a non-relativistic Hamiltonian that agrees with our effective action.

\begin{figure}[H]
\begin{center}
\includegraphics[align = c, scale=0.35]{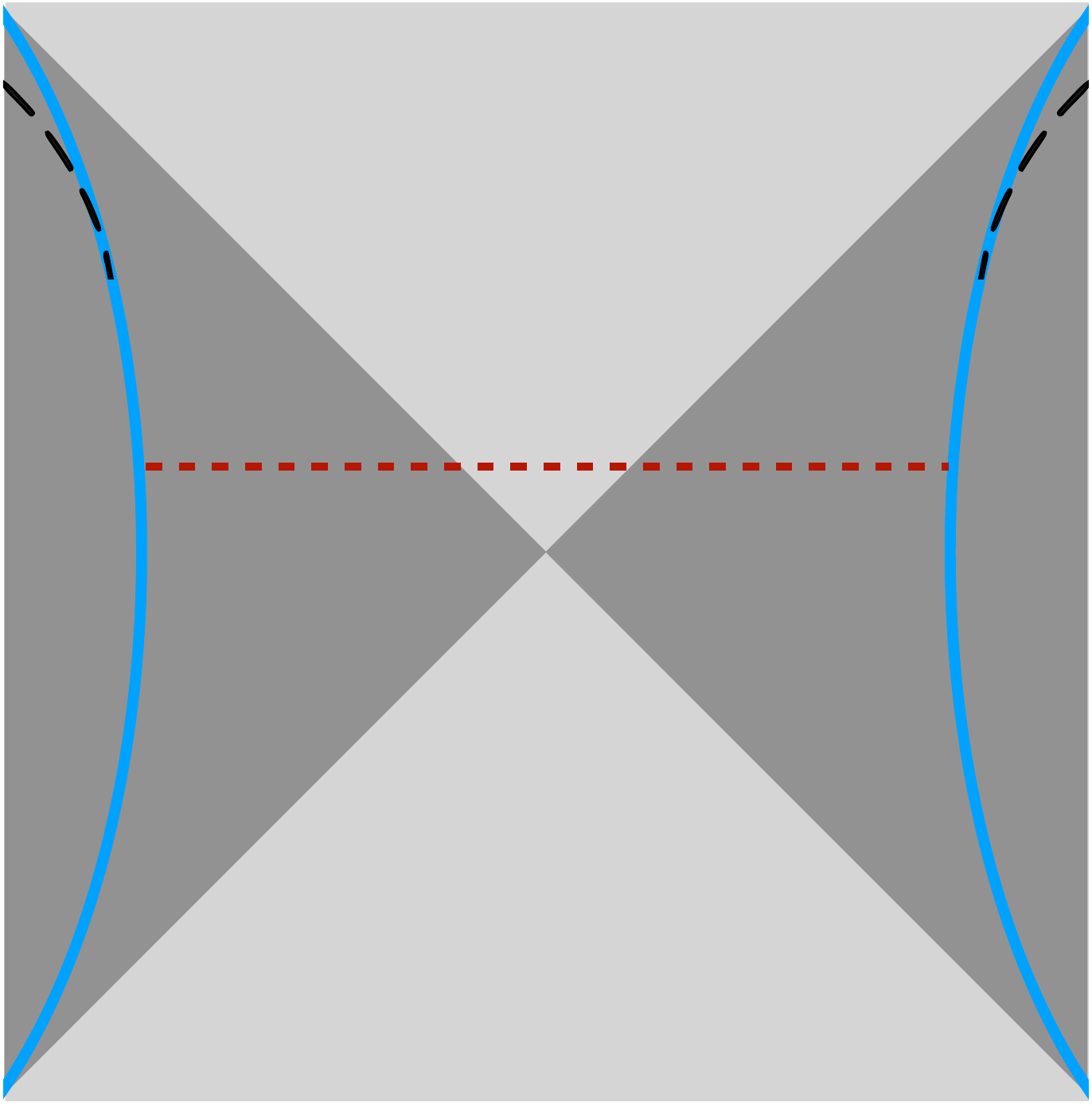}
\caption{The 2-sided, uncoupled setup. At some later time, we can insert some matter, which kicks the Schwarzian boundaries.}	
	\label{2s-kick}
\end{center}
\end{figure}

One interesting feature of this action is that the distance of closest approach (which occurs at $u=0$) is of order $\ell \sim 2\log \beta \cj$. This minimum distance should be interpreted as the complexity of formation. Since the Euclidean preparation of the state uses a strip of length $\sim \beta $ (and not $\log \beta $), we view this formula as evidence that the microscopic definition of complexity must distinguish between Euclidean and Lorentzian (unitary) evolution.

\subsection{2-sided coupled Hamiltonian \la{coupledHam}}
Now following \cite{Maldacena:2018lmt}, we consider the effect of coupling the two sides by turning on an interaction $\propto \eta O_l O_r$, where $O_l$ and $O_r$ some fields of dimension $\Delta$, evaluated on the boundary of NAdS$_2$. In the SYK model, the interaction could be generated by $\sum_i \psi^i_L \psi^i_R$.  For small enough $\eta$, we can analyze the resulting dynamics directly in the Schwarzian limit. The interaction adds a term in the effective potential
\eqn{V  = e^{2\varphi} - \eta e^{2\Delta \varphi} + {E\over N } e^\vf.  \la{eternalmatter} } 
The form of this interaction is intuitive since correlations between left and right fields decay exponentially with distance as $e^{2 \Delta \varphi}$.
We would like to ask what happens if one starts in the ground state of the coupled Hamiltonian and then adds a relatively small amount of matter. If we apply a symmetric kick on both sides, this means we start with $E = 0$ for early times and then suddenly shift the potential 
$V \to V + {E_0 \over N } e^\phi.$ at some time $\uk$.
For small $E_0/N$ the main effect is that the minimum of the potential is shifted slightly. 
This means the breathing mode of the eternal wormhole will be somewhat excited. 

\begin{figure}[H]
\begin{center}
	\includegraphics[scale=0.5]{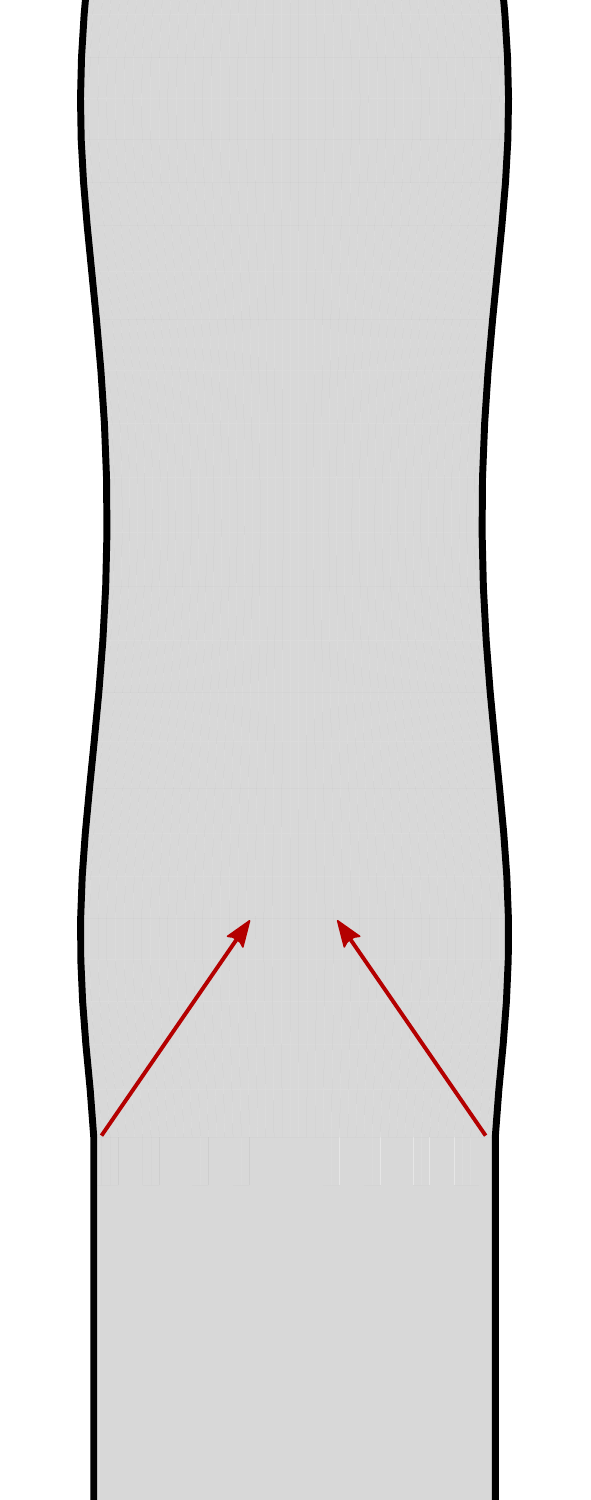} 	
	\caption{A 2-sided kick to the eternal traversable wormhole. Originally, the distance between the two sides is constant. The perturbation causes the boundaries to oscillate. This should be compared with Figure \ref{sphere}, where the complexity also oscillates after a perturbation. }
\label{diffgrav}
\end{center}
\end{figure}
%The SL(2) charges
%\eqn{\vec{Q} = m\, Y \times \dot{Y} \sim \Delta e^{2 \rho} \lp 1,0, 1 \rp}
If we kick the boundaries asymmetrically, for example by dropping a particle in from only one side, the analysis is only slightly more complicated, see Appendix A for details. The upshot is that to leading order in $1/N$ we may pretend the kick was symmetric and use the shifted potential to calculate the distance as a function of $u$.

%This means we have some non-zero $q_\pm $ in the notation of Maldacena-Qi.
%The non-zero $q_\pm$ will lead to oscillations of the boundary that are periodic with exactly the AdS frequency. But the non-zero $q_0$ terms will lead to oscillations that are periodic with a somewhat different period.

%Let us pause to discuss the near-extremal Reissner-Nordstrom interpretation of this setup.
%\eqn{V(r) = {\pd_r f^2 \over 4r^2} }
%TO DO: convert this into gloal coordinates and derive an expression for the potential barrier. I think it should look like a valley.

%The point is that as a particle oscillates in the potential, momentum conservation means that the boundary also oscillates.

\subsection{1-sided black holes \la{1s}}
\def\deow{\ell_\text{EoW}}
We may also consider a single SYK system in a pure state. For certain choices of pure states in SYK, one can argue \cite{Kourkoulou:2017zaj} that the gravitational dual is a single-sided black holes with an end-of-the-world (EoW) brane behind the horizon. For these states, a natural definition of distance $\deow$ is the maximum distance from the end of the world to the Schwarzian boundary. This is given simply in Poincare coordinates:
\eqn{ds^2 = {-dt_P^2 + dz^2\over z^2}, \quad \deow = \int {dz \over z} = - \log z/z_\text{EoW} = - \log t'_P(\ut)  }
So up to additive constants that we have dropped, $\phi \equiv \log t'_P = -\deow$.
Here we have assumed that the end-of-the-world brane is at very large $z/\ell_{AdS} \gg 1$, and that the Schwarzian boundary is at small $z \propto t'_P(\ut)$. We get
%	Writing a Lagrange multiplier for $f' = e^{\phi}$, we get \cite{}
\eqn{%S &= %N \int d\tilde{u} \hf {f''^2 \over f'^2} + {d \over d \tu} \lp f'' \over f' \rp \\
S = {N \over 2} \int d\tu\,  {\phi'}^2 - \lp 1 + {E_{P} \over N} \rp e^\phi . \la{purestate}}
	%In the last line, we have used that $\lambda$ is determined by the gauge constraint corresponding to the $f \to f+ c$ symmetry. 
	We have used that the brane carries an amount of Poincare energy $E_P^\text{brane} = N$, so the $E_P$ that appears in this formula is the Poincare charge of additional matter that could be added. Unlike in the 2-sided case, where there is a mismatch by a factor of 2, here a matter perturbation really is just shifting the angular momentum.

Note that we could also evolve by an alternative state-dependent Hamiltonian \cite{Kourkoulou:2017zaj} similar to the coupled 2-sided Hamiltonian. Then we would stabilize the effective potential. Just as the coupled 2-sided Hamiltonian can be thought of as the symmetry generator of global energy, we can think of the 1-sided coupled Hamiltonian as an approximate Poincare symmetry generator.

\subsection{The switchback effect \la{sec-switchback}}
We have already discussed adding matter shockwaves in the various gravity setups, or kicking the particle in TCG. Exciting matter fields in the bulk is analogous to changing the angular momentum. Both change the effective potential after the perturbation is applied. The analogy is most clear in the 1-sided case, where comparing the actions \nref{purestate} with \nref{cgea}, we see that adding Poincare charge is mathematically the same as adding angular momentum in TCG.

Here we will emphasize that the shift in the effective potential due to a small perturbation (from either point of view) is what governs the switchback effect. This is not surprising but we found the calculation (using the effective potential formalism) to be instructive.
Furthermore, there is a subtlety in both the Schwarzian analysis and in TCG: the magnitude of the matter charge $E_P$, or the change in $J^2$, depends on when the kick is applied. In the Schwarzian langauge, this is because a kick (in the SYK model, adding an instanteous source which for a small number of fermions) changes the physical energy by a constant amount. However, the physical energy is related to the Poincare energy $E_P$ by the conversion factor of $df/du = e^{\phi}$. This means that $E_P \sim \exp\lp {-\phi(\uk)}\rp$. Of course, after the kick, $E_P$ is conserved.

Similarly, the change in the angular momentum of the TCG particle depends on when the kick is applied. Indeed, in \tcg, the rule was that the kick is in the direction perpendicular to the velocity of the particle. This implies that the torque on the particle (measured with respect to the origin) will depend on the position and momentum of the particle. Here we will choose a slightly different (but qualitatively similar) rule: the perturbation is always an angular kick, with a magnitude chosen so that the energy of the TCG particle changes by a tiny fixed amount.

We would like to emphasize that the match is enough to reproduce the switchback effect. The switchback effect occurs when we go back in time, apply a kick, and then come back to the present. We then compare the resulting trajectory of the Schwarzian boundary/TCG particle with the counterfactual trajectory where no kick was applied. 

%In fact the exponential potential gives us an intuitive picture of this effect. In particular, notice that we can write the change in potential after the kick as
%\eqn{V = e^\phi \to e^{\tilde{\phi}}, \quad \tilde{\phi} =  \phi + \log \lp 1 + E_P/N \rp \\}
%We see that the effect of the shockwave is to simply shift $\tilde{\phi}$ by a constant amount, without changing the potential $V(\tilde{\phi})$ or the velocity.
%Now let us neglect the change in the energy of the particle due to the kick. We can then solve for the trajectory. The only effect is to adjust the zero mode $u_0$.
%\eqn{ \phi_0(u_s)-\phi_c = \phi_0(u_s - u_0) }
%\eqn{u_0 = \int {d\phi \over \sqrt{2(E - V)}} \approx {\log (1 + E_P/N) \over \sqrt{ 1 - e^{\phi}} } } 
%This means to solve for $\phi(u)$ in the presence of a shock, we simply 

%If $|\uk|$ is sufficiently large, then the effective potential gives an intuitive picture of what happens. Since $E_P \propto \exp\lp {-\phi(\uk)}\rp$, this term in the effective potential grows as a function of $|\uk|$. Around the scrambling time, $\phi \sim \log N$ and the term $E_P$.

%For small $|\uk|$, we expect that the difference in the complexity starts to grow with the Lyapunov exponent $2\pi/\beta$ until the scrambling time, after which it grows linearly.

\def\us{u_\text{switch}}

We also consider the same effect in the two-sided case, see Appendix \ref{appendixBoost}. For the two-sided case, inserting matter is not precisely the same as adjusting the angular momentum. For the former, the potential shifts by $\sim e^\phi$ whereas in the latter, we would get a shift by $\sim e^{2\phi}$.

The equations of motion for this particle is $\ddot{\rho} = 4J^2 e^{-2\rho}$. The solution is
\eqn{ e^{2\rho} = {4 J^2 \cosh^2 \lp   \pi u /\beta \rp \over  (\pi /\beta)^2 }, \quad u \le \uk  \\
e^{2\rho} = {4 (J')^2 \cosh^2 \lp  v (u-u_0) \rp \over v^2 }, \quad u > \uk }
We match the two solutions at $u = \uk$. We parameterize the kick in terms of the change in energy $\epsilon$ of the TCG particle:
\eqn{J'^2 = J^2 + {\epsilon \over 2} e^{2\rho} = J^2 \lp 1+ 2  \epsilon \lp \pi \over \beta \rp^2 \cosh^2(\pi u/\beta) \rp. } 
The asymptotic speed $v$ is set by $ v^2 = (\pi/\beta)^2+ 2 \epsilon$. %Then, at time 0,
%\eqn{e^{2 \rho} = \frac{\cosh ^2\left(\sqrt{2 \epsilon +1} u_s+\cosh ^{-1}\left(\cosh \left(u_s\right) \sqrt{\frac{2 \epsilon +1}{\epsilon  \cosh \left(2 u_s\right)+\epsilon +1}}\right)\right)}{2 \epsilon +1} }
%At time $\us$ the value of $J$ changes by an amount of order $1/N$. 
%We will solve for the new trajectory in perturbation theory $\rho = \rho_0 + \delta \rho$. 
%\eqn{ \delta \ddot{\rho} =  8  J_0^2  e^{-2\rho_0}  \lp    {\delta J \over J_0} - \delta \rho \rp  \approx \lp 2\pi \over \beta \rp^2 \lp {1 \over N} -2 \delta \rho   \rp}
%Now we can use this to easily calculate the speed at time $-\us$.
%\eqn{ \hf v^2 = E - J^2 e^{-2 \rho}  = {2\pi^2 \over \beta^2} \lp 1 - {1 \over \cosh^2(u \pi/\beta) }\rp .}
%\eqn{\Delta C= \int v \, du = \int \sqrt{E- e^{-\rh} } }
%The speed at time $-u$ is $v = \int a du = $
%\eqn{\delta \ddot{\rho}/2 =  L^2 e^{-2\rho_0} \delta \rho +  L \delta L e^{-2\rho_0} }
%Since angular momentum is conjugate to $\theta$, we are really asking
%\eqn{\delta \rho =  \{\rho(0), \theta(-u) \} }
%Imposing the initial conditions $\delta \rho = \delta \dot{\rho} = 0$, we get
Evaluating the solution at $u=0$ and expanding to first order in $\epsilon$ gives 
\eqn{ \delta \rho = \epsilon \lp \beta  \over \pi \rp^2 \sinh^2 \lp  \pi u \over \beta \rp. \la{switchback}}
Of course, the same calculation also applies in JT gravity. Translating variables and units, we can work out the effect of a perturbation (as measured by the SYK Hamiltonian, perturbing a fermion adds energy $\delta H_\text{SYK} = 2 \Delta \cj$): 
\eqn{ -\delta \phi = {\cj \delta H_\text{SYK} \over 4 N \alpha_s} \lp \beta  \over \pi \rp^2 \sinh^2 \lp  \pi \uk \over \beta \rp \la{switchback}}

Finally, well before the scrambling time, it is natural to assume that the complexity of this pure state is well-approximated by thermal size.
We can calculate the thermal size either directly in the large-$q$ SYK \cite{Qi:2018bje}, or using the bulk dual/Schwarzian description \cite{Lin:2019qwu}, which is in some linear approximation is basically the 2-sided length. Copying equation (6.111) \cite{Lin:2019qwu} in our notation:
\eqn{ \text{size} = \Delta  { \cj \delta H_\text{SYK} \over  4 \alpha_s} \lp \beta  \over \pi \rp^2 \sinh^2 \lp \pi \uk \over \beta\rp }
This is consistent with the relation $\sum_i \psi_L^i \psi_R^i \sim N e^{\Delta \phi}$.
%Both methods yield results in agreement with \ref{switchback}.
We view the agreement between these formulas obtained from $\sim 3.5$ different perspectives as a somewhat nontrivial consistency check between existing ideas about complexity, size, and bulk gravity.

\subsection{Beyond low temperature: large $q$ SYK}
Here we comment on the SYK model in the limit where $q \to \infty$, $N \to \infty$, and $q^2/N \to 0$, but at finite temperature  $\beta \cj \sim O(1)$. In this limit, we need to go beyond the Schwarzian approximation and use the master field variables (see, e.g., \cite{Maldacena:2018lmt}) which at large-$q$ are $g_{LL}(u_1,u_2)$ and $g_{LR}(u_1,u_2)$. Focusing on the left-right correlator, % For our purposes, we only need the second one:
\eqn{%{1 \over N} \sum_i \ev{\psi_L^i(u_1) \psi_L^i(u_2)} =  G_{LL} = G^0_{LL} e^{g_{LL}/q},\\
	% \quad 
{1 \over N} \sum_i \ev{\psi_L(u_1)^i \psi_R(u_2)^i} = G_{LR} = G^0_{LR} e^{g_{LR}/q}. }
%In the following discussion, we will suppress the $G_{RR}$ correlator, which appears identically to $G_{LL}$. %The large $N$, large $q$ effective action is 
%\eqn{S[g_{LL}, g_{LR}] = {N \over q^2} \int du_1 du_2 {1 \over 4} \lp \pd_1 g_{LL} \pd_2 g_{LL} - \pd_1 g_{LR} \pd_2 g_{LR} \rp - \cj^2 \lp e^{g_{LL}} + e^{g_{LR}} \rp.}
Assuming that $g_{LR}(u_1,u_2) = g_{LR}(u_1-u_2)$ on shell, the large-$q$ equations of motion give \cite{Maldacena:2016hyu}:
\eqn{\pd_u^2 g_{LR} =  - 2 \cj^2 e^{g_{LR}}.}
Given that the precise bulk dual of SYK is not known at finite temperature, we do not have a precise expression for the distance between the two sides. Nevertheless, it is reasonable to think that correlation functions decay exponentially with distance $G^0_{LR} e^{-\Delta d_{LR}} = G_{LR} $. This suggests that we can match $d = -g_{LR} + \text{const}$ in the semi-classical limit. With such an identification, we see that even at finite temperatures, the distance behaves like a non-relativistic particle in an exponential potential. Similarly, if we couple the two sides, the potential is modified in the same manner as in the Schwarzian limit.

{\bf Tomperature versus temperature.}\\
For a generic $q$-local chaotic quantum system, we expect the complexity to obey
\def\ct{\mathcal{T} }
\eqn{dC/du =\ct S.}
For an infinite dimensional quantum system like a quantum field theory, it is natural to identify the tomperature $\ct$ with the temperature $T$. This however cannot be correct if we are dealing with a system with a finite dimensional Hilbert space like in SYK. In particular, even as $T \to \infty$, the growth of the state complexity cannot be faster than the growth of the complexity of the unitary that implements time evolution, which stays finite.

For times that are not exponentially long, we expect that simple 2-pt functions should behave like $ G_{LR} \sim \exp \lp \Delta C/S\rp$, where $\Delta$ is the dimension of the operator. (For a single SYK fermion, $\Delta = 1/q$). Hence we define the tomperature
\eqn{ \ct = -{1 \over \Delta} {d \over du} \log G_{LR}. }
In the large-$q$ model, this is exactly the $\ct = \lambda_L/(2\pi) $.  This reduces to $\ct \approx T$ at low temperature. Note here that $\lambda_L$ is the chaos exponent defined by the 4-pt function, whereas the tomperature is defined by the 2-pt function. This raises the interesting question of whether the large-$q$ SYK model could still be (in some unorthodox sense) maximally chaotic at finite temperatures.

\section{Discussion}
We end with some speculative remarks and open questions.

\begin{enumerate}
	\item The applicability of these ideas in global AdS$_2$ suggests that size and complexity are intimately related to the emergence of gravity, even in situations where there are no black hole horizons. This seems to be an advantage over Erik Verlinde's ideas \cite{Verlinde:2010hp}.

	\item We have focused on the complexity = volume conjecture. One can wonder about the complexity = action conjecture. The results of \cite{Brown:2018bms, Goto:2018iay} show that one needs to go beyond JT gravity (e.g., know something about its higher dimensional origins) in order to get a reasonable Wheeler-de-Witt action in two dimensions. On the other hand, the effective dynamics of volume (which in two dimensions is just distance) exists entirely within the Schwarzian approximation. On effective field theory grounds, the Schwarzian should arise whenever there is an emergent near-conformal symmetry \cite{Maldacena:2016upp}. So the dynamics of volume discussed here is in some sense universal, whereas the dynamics of Wheeler-de-Witt action could be quite model-specific.
		
	\item What does this teach us about complexity? If we take complexity = volume seriously, then for systems dual to NAdS$_2$ gravity, complexity must have a much simpler effective description at large $N$ and low temperatures. Such an extraodinary simplification is remarkable, given that Nielsen-like approaches to complexity geometry feature a highly anisotropic metric. Attempts to analyze the geometry are not easy at small $N$ \cite{Brown:2019whu}, and seem to get harder (not easier) at larger $N$. Clearly we are not understanding the best way to define or analyze complexity.

%	\item What does this teach us about the ``right'' definition of complexity for SYK, e.g., one that would match bulk properties? In order to reproduce the effective potentials discussed in this paper, it seems likely that a smooth, low-curvature notion of complexity is needed. It is  hard to imagine that a Nielsen-like definition of complexity geometry could do the job, since the cut locus in the Nielsen geometry is so close to the origin. 
%Furthermore, the right definition of complexity should somehow have the property that there is a much simpler effective description when $N\to \infty$. One prototype of such a simplification is when a geometry has $SO(n)$ symmetry for large $n$. If we are only interested in the distance from the origin, we may exchange the gazillions of Euler angles for an angular momentum barrier. However, the full complexity geometry is expected to be highly anisotropic, so something less trivial must be going on.

	\item On a related note, we are matching complexity geometry to JT gravity calculations. But the dual of JT gravity is really a disorder average of an ensemble of quantum mechanical systems \cite{Saad:2019lba}. One might wonder what the proper role of the disorder average is in a good definition of complexity geometry. Perhaps complexity is simpler to define or calculate when we consider an ensemble of Hamiltonians. An example of a toy model of complexity where a disorder average leads to significant simplification is \cite{Lin:2018cbk}. %The simplificty of the Schwarzian dynamics of the volume cries out f9to understand how such a fine-grained quantity like complexity could behave so simply.

	\item Lastly, note that the dynamics of length in JT gravity at low temperatures only depend on the matter via the SL(2) charges. This is a form of the equivalence principle. Since the length is related to size and complexity, this implies that both of these quantities also satisfy the equivalence principle. For example, the size of an operator in SYK should not depend on which fermion $\psi_i$ we are considering; more generally the size of composite operators like $\psi_i \pd^n \psi_i$ should only depend on the dimensions of these operators. It would be interesting to understand the microscopic origin of the equivalence principle, which again points to a somewhat surprising simplicity of these fine-grained quantities.
		
%	\item One odd feature of complexity geometry is that we are matching a length in a Euclidean geometry to a length in a Lorentzian geometry. Furthermore, the particle on the complexity geometry is entirely classical, whereas the Schwarzian boundary particle is subject to quantum fluctuations. 

		%From the Nielsen complexity geometry point of view, there is no obvious simplification. Another way of asking this question is the following. From the Nielsen definition (or the Brown-Susskind modification), the complexity geometry is highly anisotropic. But an effective potential suggests that we can ignore this in many situations. Can this be made more precise?
	%\item We seem to be able to explain the gravitational force in empty AdS. This seems to be one key difference over the ideas of E Verlinde
\end{enumerate}

\section*{Acknowledgements}
We thank Adam Brown, Daniel Harlow, Juan Maldacena, Douglas Stanford, and Ying Zhao for discussions. H.L. is supported by an NDSEG fellowship. L.S. is supported by NSF Award Number 1316699.

\mciteSetMidEndSepPunct{}{\ifmciteBstWouldAddEndPunct.\else\fi}{\relax}
\bibliographystyle{utphys}
\bibliography{BigReferencesFile.bib}{}

\appendix

\section{Geodesic length in the Schwarzian formalism}
First let us discuss the distance between the 2-sides in the Schwarzian limit. 
The distance between two points (in AdS units) on opposite boundaries of NAdS$_2$ can be easily calculated in the embedding space formalism:
\eqn{\dlr = \cosh\inv \lp X_L \cdot X_R\rp \approx - \log \left[ T'_L T'_R\over \cos^2\lp {T_L - T_R\over 2}\rp \rb .}
In the last equality, we have subtracted off a divergent term.
Now suppose we can find a gauge where $T_L(\tu) = T_R(\tu)$ for all $\tu$. For these states, define $T'(\tu) = e^\vf$. This new variable $\vf$ is essentially the geodesic length 
\eqn{\dlr + \text{const} = -2\varphi.} We will now derive a very simple effective action for $\vf$. 
We start with the Schwarzian action in the 2-sided case. In the SYK model, the coefficient of the Schwarzian action is $N \alpha_s/\cj$, where $\alpha_s \simeq 1/(4q^2)$ in the large-$q$ limit; we have defined the rescaled time $\tilde{u}$ to absorb the factors of $\cj \alpha_s$.
% Note that $du/u^2 = 1/u = \cj /\tu$, which cancels the factor of 
\eqn{S &=  \text{Sch}(T_L) +  \text{Sch}(T_R) \\% {N \alpha_s } \int du \{\tan T_L(u)/2 \} + \{\tan T_R(u)/2 \}.\\
	\text{Sch}(T) &=  N \int d\tu \, \frac{T''^2-T'^4}{2 T'^2} - {d \over d\tu} \lp {T'' \over T'} \rp
\la{sch}}
Dropping the total derivative and adding a Lagrange multiplier to enforce $T'(\tu) =  e^\varphi$,
	\eqn{S &= N \int d\tu  \lp \frac{T''^2-T'^4}{ T'^2}\rp - \lambda (T' -e^{\vf}  )\\
	       &= N \int d\tu \, {\vf'}^2 - e^{2 \varphi}  - \lambda (T' -  e^{\vf}) . }
The equations of motion for $T'$ set $\lambda ' = 0$. To determine the constant value of $\lambda$, we need to remember the gauge constraint. Since an overall translation $T \to T + c$ to both the boundary and matter is a gauge symmetry, we need to set the corresponding generator to zero\footnote{For configurations where $T_L(u) = T_R(u)$ the other SL(2) charges corresponding to an overall momentum and boost are automatically zero.}. The Noether procedure then gives
		\eqn{0 &= {\pd L \over \pd T'}  -{d \over d \tu}  \pa{L}{T''}= \lambda + 2  \lp T' + {{T''}^2 \over T'^3} -{T''' \over T'^2}\rp \\
	       &= \lambda  - E,  }
where $E$ is the matter global energy charge. This gives 
			\eqn{S  &= N \int d\tu \, {\varphi'}^2 -  e^{2 \varphi}  - {E  \over N} e^{\vf} . }

This derivation is easily generalized to the other cases we consider in the paper. The key point we would like to emphasize is that the value of the Lagrange multiplier is the SL(2) matter charge that corresponds to the time variable we are using. So for example, if we use Poincare time $f$ as in section \ref{1s}, the value of the Lagrange multiplier $\lambda (f' - e^\phi)$ will be determined by the SL(2) charge associated to the Poincare symmetry $f \to f+c$.

\section{Momentum and length in the eternal traversable wormhole}
\la{appendixEternal}
In this appendix, we explore how the length and momenta change when we drop in a single fermion. 
\eqn{\vec{Q} = (B, P, E) \sim \Delta \beta \cj \lp 1,0, 1 \rp}
Note that the definitions of charges $Q$ are dimensionless, to make them dimensionful we use a factor of $\beta$. This implies that dropping in a fermion changes the boundary energy $\delta \ev{H_R} \sim \cj$.

We start in the ground state where $T_L = T_R = T' \tilde{u}$, and then consider a small fluctuation in the boundary positions. It is convenient to decompose into symmetric and anti-symmetric parts: 
\eqn{ T_L = T' \tilde{u} + \chi(\tilde{u}) + \psi( \tilde{u} ), \quad T_R = T' \tilde{u} + \chi(\tilde{u} )-\psi(\ut ).} SL(2) charge conservation gives us
			\eqn{\psi &= -{B \over  8 N T' } \lp T \cos T  \rp\\
				\lp \pd_T^2 + \omega^2 \rp \chi'  &= {E \over 2NT'} \RA \chi = {E \over 2 N \omega^2 T'}\lp1 - \cos(\omega T) \rp .
}
Note that $B$ and $E$ go from $0$ to some finite value at $t=0$.
We should enforce continuity of $\chi, \psi$ and their first derivatives in order that the boundary is continuous when matter is inserted. Then integrating the first equation over an infinitesimal time interval near $t=0$ gives us continuity of $\chi^{(3)}$.
Then the length between the two sides
\eqn{\delta \ell \sim - \delta \log \left[ T'_L T'_R \over  \cos^2 \lp {T_L - T_R \over 2}\rp \right] = -  2 \chi' + O\lp {1 \over N^2}\rp, }
So to leading order in $1/N$, we find the distance
\eqn{\dlr =  \ell_0 + {E \over NT' \omega^2}  \sin \omega T.}
Note that we could have guessed an expression of this form by naively ignoring the $P,B$ charges and using the effective action given by \ref{eternalmatter}.
It might seem rather peculiar that the oscillations are at a frequency $\omega$ which is determined by the breathing mode and not the AdS frequency, 
especially since there is a close connection between length and momentum \cite{Lin:2019qwu}. However the momentum is given by
\eqn{P \propto (\pd_L - \pd_R) \dlr \approx 2\lp \psi +\psi''\rp  \sim \hf B \sin T .}
So there is no contradiction as long as we remember the relative signs. Note that in the usual coupled wormhole, the breathing mode and the AdS frequency are of the same order of magnitude, but in a slightly more complicated model, they can be different \cite{Lin:2019qwu}. 
%Notice that the periods are incomensurate. The question is which one of these effects is related to complexity. 
%\eqn{E/N = - 2 T' d_t \lp d_t^2 + 2(1-\Delta) \rp \chi, \quad P/N = 2T' }
%Consider a non-relativistic particle in Euclidean space. We have the position operators and the angular momenta:

\section{Effective dynamics for the boost $H_L - H_R$ \la{appendixBoost}}
%In a 2-sided setup, one usually does not consider evolution by $H_L - H_R$. 
%The reason is that for the thermofield double state $(H_L - H_R) \ket{\text{TFD}} =0$. This is very special to the TFD and means that the left and right horizons meet at a point (the bifurcate horizon). 
%In a more typical state, for example one with some matter charge $E_0$, the horizons will not meet. This means that the distance between the two sides is not invariant under $H_L - H_R$. 
Here we will consider the effective dynamics of the length under evolution by a boost $H_L - H_R$. If the state we are interested in is exactly the TFD, the length of course does not evolve since $(H_L - H_R) \ket{\text{TFD}} = 0$. However, even a tiny perturbation to the TFD will cause the length to start growing. This is just another aspect of the butterfly effect. Here we will see that this aspect of chaos is just the instability of an inverted harmonic oscillator.

To work out the effective action of $H_L - H_R$, it is convenient to go to Rindler coordinates:
\eqn{ds^2 = d\rho_R^2 - \sinh^2 \rho_R\, d\tau^2. \la{rindler}}
We will be using the convention that $\tau$ runs up on the left and down on the right. With this convention, we will consider trajectories where $\tau_L(\tu) = \tau_R(\tu) = \tau(\tu)$, where $\tu$ now means evolution with respect to $H_L - H_R$. 
It is convenient to define $\tau' = e^{ \rho}$, where $\rho$ is up to a constant shift equivalent to the Rindler radius $\rho_R$ that appears in \nref{rindler}. The Schwarzian action then becomes
\eqn{S= {N} \int d \tilde{u} \, {\tau''^2+ \tau'^4 \over  \tau'^2 } \RA
S[\rho] = {N } \int d \tilde{u}\, {{\rho '}^2} +  e^{2 \rho} -  B  e^{\rho}. }
Here $B$ is the difference of the boost charges $B = Q_L^B - Q_R^B$. The minus sign comes from fact that we are evolving by $H_L - H_R$.
For the TFD, we get $B =  {4\pi / \beta }  $. A solution of this effective action is simply $\tau' = 2 \pi/\beta$. %; this is approximately true for any states that are approximately thermal.

Notice that the effective potential has a maximum instead of a minimum --- this instability may be viewed as the origin of chaos! If we can start with the TFD and add a small amount of matter (which changes $B$) the particle will not be exactly at the top of the hill. Its length will start to quadratically grow at first, and then transition to an exponential until the scrambling time.

We can think of this as giving an effective action for the distance from the boundary to the point $\rho = 0$. 
The point $\rho = 0$ can be defined in a gauge-invariant way by saying that it is the point that would be the bifurcate horizon if no matter was added. 
This is a rather awkward counterfactual and teleological notion; the only advantage of such a definition is that it also makes sense in a 1-sided context. When there are 2 sides, it is much simpler to think of this as an effective action for the distance between the two sides, but with time evolution defined by $H_L - H_R$ instead of $H_L + H_R$.

As an application of this formalism, we derive the Qi-Streicher formula. First, we note that $P \sim dS/du \sim dL/du$. So up to a constant, we can calculate the size by calculating the geodesic length. Furthermore, if we are interested in times before the scrambling time, we may linearize
\eqn{{d^2 \over du^2} \delta \rho = \lp {2 \pi \over \beta}\rp^2  \lp \delta \rho  - {\epsilon }\rp }
Here $\epsilon = \beta \delta B/(2\pi) $ which for a single fermion is of order $\epsilon \sim \cj \beta^2 (\cj/N\alpha_s)$. The solution to this equation is a sum of exponentials. The boundary condition then imposes
\eqn{\delta \rho \propto  { \lp \beta \cj \rp^2 \over N}  \sinh^2 \lp \pi u \over \beta \rp }
Here we have ignored the $1/N$ change to the energy that results when we add a particle. This is essentially the same formula that we found in \ref{switchback}; from this point of view, chaos is just a manifestation of a particle rolling near the top of a potential well.

\section{Another toy model}
Here is an example of a complexity metric which could match the asymptotic properties of the 2-sided Schwarzian action
\eqn{ ds^2 = dr^2 + \sinh^2\lp r \over l\rp d\theta^2 + \cosh^2\lp {r \over 2l} \rp d \phi^2. \la{toy2} }
This can be viewed as the Poincare disk with a compact extra dimension that grows exponentially with radius. At large radius, we get a particle in an effective potential
\eqn{V \sim J_\theta^2 e^{-2r/l} + e^{-r/l} J_\phi }
This space is negatively curved
\eqn{R \, l^2 = -{7 \over 2  } + {1 \over 1 + \cosh (r/l)} }
We see that the curvature in units of $1/l^2$ varies from $-7/2$ near $r=0$ to $-3$ as $r \to \infty$. Unlike the hyperbolic plane, this space is not homogeneous. Note that if TCG arises as a 2-dimensional surface embedded in the full complexity geometry according to the rules proposed in \cite{Brown:2017jil}, there is no particular reason why it should be homogeneous.

Another example of a geometry with exponential potentials is a Euclidean version of de Sitter space
\eqn{ ds^2 = dt^2 + \sinh^2 t \, d\theta^2 + \cosh^2 t \, d \phi^2.}
The effective potential in this geometry has two charges that are separately conserved $L_\theta$ and $L_\phi$. Unlike \nref{toy2}, this geometry is homogeneous. We are engaging in what might be called ``complexity phenomenology,'' e.g., we are trying to make reasonable models of complexity geometry that are consistent with some principles and fit certain ``data.'' A derivation of such geometries starting from a principled definition of complexity would obviously be more satisfactory.

\end{document}